\documentclass[preprint]{ptephy_v1}%%%%%% to generate preprint number

\preprintnumber{KUNS-2925} %%% %%% Insert preprint number here

\usepackage{color}

\usepackage{amsmath}
\usepackage{amssymb}
\usepackage{amsfonts}
\usepackage{amsthm}
\usepackage{bm}
\usepackage{hyperref}
\usepackage{url}
\hypersetup{colorlinks = true, allcolors = blue}

\DeclareMathOperator*{\Tr}{Tr}

\begin{document}
\title{Structural analysis of a collective subspace in the dynamical generator coordinate method}

\author{Norihiro Hizawa}
\affil{Department of Physics, Kyoto University, Kyoto 606-8502, Japan \email{hizawa@ruby.scphys.kyoto-u.ac.jp}}

\begin{abstract}
In nuclear theory, the generator coordinate method (GCM), a type of configuration mixing method, is often used for the microscopic description of collective motions.
However, the GCM has a problem that a structure of the collective subspace, which is the Hilbert space spanned by the configurations, is not generally understood.
In this paper, I investigate the structure of the collective subspace in the dynamical GCM (DGCM), an improved version of the GCM.
I then show that it is restricted to a specific form that combines tensor products and direct sums under reasonable conditions.
By imposing additional specific conditions that are feasible in actual numerical calculations, it is possible to write the collective subspace as a simple tensor product of the collective part and the others.
These discussions are not dependent on the details of the function space used for generating the configurations and can be applied to various methods, including the mean-field theory.
Moreover, this analytical technique can also be applied to a variation after projection method (VAP), then which reveals that under a specific condition, the function space of the VAP has an untwisted structure.
These consequences can provide powerful tools for discussing the collective motions with the DGCM or the GCM.

\end{abstract}

\subjectindex{D10}
\maketitle

\section{Introduction}
In strongly correlated systems, emerging degrees of freedom due to quantum superposition effects can play an essential role in the system's dynamics.
Such degrees of freedom are often cooperative in terms of a single particle, and the phenomena associated with them are referred to as {\it collective motions} \cite{Ring_Schuck}.
These, including nuclear fission and shape oscillations, are often well understood qualitatively and phenomenologically.
On the other hand, describing the collective motions based on microscopic degrees of freedom is one of the most challenging tasks in nuclear theory.

In principle, all information on the collective motions should be contained in a microscopic Hamiltonian.
However, dealing with the eigenvalue problem exactly is often difficult, even for numerical cases.
In addition, the fact that the general collective motion is not a rigorously defined object makes the problem more complicated.
Namely, even if one succeeds in diagonalizing the Hamiltonian, how to extract the collective motions is highly nontrivial.
In particular, the interaction between nucleons is not completely determined in nuclear theory, and then effective interactions such as Gogny \cite{Gogny} and Skyrme \cite{Skyrme} are often used in the microscopic calculations.
The time-dependent density functional theory (TDDFT) \cite{RG84} describes the exact time evolution of a quantum many-body system under a given initial condition in principle, which is one of the leading methods to describe the collective motion.
However,  the TDDFT loses information on the wave function itself, which can be disadvantageous to the theoretical understanding of the collective motion.
Therefore, a universal and powerful approximation method is necessary for investigating such complex collective phenomena with many uncertainties.
The mean-field theory is one of the significant approximation methods for analyzing quantum many-body systems.
In this theory, states are approximated by the wave functions of the independent particle approximation (IPA) type.
However, according to the Wick theorem, all correlation functions can be written down in polynomials of two-point correlation functions.
This fact implies that the naive mean-field theory is incompatible with the collective motion.

A generator coordinate method (GCM) \cite{HW53, Ring_Schuck} is one of the beyond mean-field methods.
The GCM is a method of mixing configurations along a {\it path} specified by generator coordinates, and it is considered that taking an appropriate path enables the incorporation of the quantum correlations necessary for the description of the collective motion.
Numerical calculations of the GCM already started decades ago \cite{FV76, BD90} and have been actively performed in recent years with the development of computer technology \cite{Bender2003, NV11,E16,RR18,SO06,BH08,YM10,TR11,RE11,YB13,FS13,BA14,Yao2014,YZ15,EJ20}.

The most severe problem of the GCM is that a method to determine the appropriate path for describing a collective motion of interest is nontrivial, and an empirical method is often used in current nuclear physics.
However, it is known that the method gives an incorrect moment of inertia, even for the translational motion, which is the simplest collective motion.
Peierls and Thouless improved this empirical path in terms of symmetry and gave proper descriptions of the translational motion and the rotational motion \cite{PT62}.
This approach is referred to as a double projection method, and its numerical applications have recently been performed \cite{BR15,EB16, ST15, Shimada2016, Ushitani2019}.
For more general collective motions, Goeke and Reinhard improved the empirical GCM in terms of conjugate momentum.
This method is referred to as a dynamical GCM (DGCM) \cite{GR80}.
Their discussion is based on a type of differential expansion called a Gaussian overlap approximation (GOA) \cite{Ring_Schuck}, and the DGCM yields an appropriate moment of inertia within the GOA.
A self-consistent collective coordinate method (SCC) is similar to the DGCM in that it considers the effect of the conjugate momentum \cite{MM80, M84, MN00, NM06}.
However, note that the degree of freedom in the SCC is the path itself, and it has a different theoretical framework from the DGCM (GCM), which deals with the mixture of the path.
In recent years, an application of the DGCM based on the GOA \cite{KE08} and direct applications for a particle number \cite{HH21} and a quadrupole oscillation \cite{HH22} have been performed.

Since the DGCM (GCM) is a kind of configuration mixing method, its essential degree of freedom is not the path but the Hilbert space spanned by the path, namely a {\it collective subspace}.
Thus, its validity should be evaluated with the structure of the collective subspace.
However, since the GCM has the inevitable ambiguity, it is generally difficult to identify the structure.
This problem is partially carried over to the DGCM, so it is difficult to discuss it without any assumptions in the DGCM.
Thus, the primary purpose of this paper is to clarify the structure of the collective subspace in the DGCM under several plausible conditions.
Toward this purpose, I analyze it in terms of a tensor product in Hilbert spaces.

The paper is organized as follows.
In Sec. II, I give a brief review and some comments on the GCM and the DGCM.
In Sec. III, I consider the simple collective motion where there is no interaction with the internal degrees of freedom and discuss how it can be described in the framework of the DGCM.
I also show an application of this technique to the variation after projection method (VAP).
The summary and future perspectives are discussed in Sec. IV.

\section{A brief review and some comments on the GCM and the DGCM}
\subsection{GCM}
At first, I briefly review the generator coordinate method (GCM) \cite{HW53}.
The GCM is a kind of configuration mixing method with many-body wave functions $\{|q\rangle \}_q$, where the continuous parameter $q$ is referred to as a generator coordinate.
Then, a given Hamiltonian $\hat{H}$ is diagonalized in the subspace $\mathcal{H}_{C}$ spanned by the GCM basis functions $\{|q\rangle \}_q$,
\begin{equation}
  \mathcal{H}_{\mathrm{GCM}} := \overline{\mathrm{Span}}(\{|q\rangle \,|\,q\in \mathcal{M}_{C}\}),
\end{equation}
where $\overline{\mathrm{Span}}$ is a closed linear span.
Note that the parameter manifold $\mathcal{M}_{C}$ is a set of $q$ and often omitted as $\{|q\rangle\}_q$ in this paper.
Although complex numbers can be considered as the generator coordinates  \cite{Ring_Schuck, jan64, BW68}, all coordinates are assumed to be real in this paper.
In addition, one can choose the dimension of $q$ arbitrarily, but since it is not essential in the following discussion, a one-dimensional coordinate $q$ is adopted in this paper.
Thus, $\mathcal{M}_{C}$ is a one-dimensional real manifold.
The path $|q\rangle$ can be chosen arbitrarily from a total Hilbert space, but for practical purposes, it is essentially restricted to within some function space.
For example, the IPA wave function \cite{Ring_Schuck, FV76} or an AMD wave function \cite{KK12} is often used in nuclear theory.
Notice that $\{| q \rangle\}_q$ do not have to be orthogonal to each other.

Any state contained in the collective subspace $\mathcal{H}_{\mathrm{GCM}}$ can be represented as
\begin{equation}
  \label{wfgcm}
  |\Psi\rangle_{\mathrm{GCM}}=\int dq\,f(q)|q\rangle,
\end{equation}
where $f(q)$ is a weight function.
Then, the eigenvalue problem of $\hat{H}$ on the collective subspace is formulated as
\begin{equation}
  \hat{\pi}_{\mathrm{GCM}}\left(
  \hat{H}|\Psi\rangle_{\mathrm{GCM}} - E|\Psi\rangle_{\mathrm{GCM}}
  \right) = 0,
\end{equation}
where $E$ is an eigenenergy and $\hat{\pi}_{\mathrm{GCM}}$ is projection onto $\mathcal{H}_{\mathrm{GCM}}$.
This equation is equivalent to
\begin{equation}
  \langle \Psi'|\left(
  \hat{H}|\Psi\rangle_{\mathrm{GCM}} - E|\Psi\rangle_{\mathrm{GCM}}
  \right) = 0
  \quad \mathrm{for \ any \ } |\Psi'\rangle \in \mathcal{H}_{\mathrm{GCM}}.
\end{equation}
Therefore a generalized eigenvalue problem is obtained as
\begin{equation}
    \label{eq:HW_GCM}
  \int dq\left[\langle q'|\hat{H}|q\rangle - E\langle q'|q\rangle\right]f(q)=0.
\end{equation}
In a context of the GCM, this is called a Hill-Wheeler equation \cite{HW53}, and $H(q', q) := \langle q'|\hat{H}|q\rangle$ and $I(q', q) := \langle q'|q\rangle$ are referred to as a Hamiltonian kernel and a norm (overlap) kernel, respectively.
Then, the eigenenergy can be evaluated as
\begin{equation}
  \label{eq:E}
  E = \frac{{}_{\mathrm{GCM}}\langle \Psi|\hat{H}|\Psi\rangle_{\mathrm{GCM}}}
  {{}_{\mathrm{GCM}}\langle \Psi|\Psi\rangle_{\mathrm{GCM}}}.
\end{equation}

In case $\{|q\rangle\}_q$ are not linearly independent, the problem of overcompleteness arises \cite{Ring_Schuck}.
If there is a linearly dependent mode defined as $\int dq\, f(q)|q\rangle = 0$, this $f$ satisfies the Hill-Wheeler equation for any $E$.
Then, the GCM has a solution with the undefined eigenenergy, and Eq. (\ref{eq:E}) is not defined.
The problem can be eliminated by removing these modes when solving the Hill-Wheeler equation.
In the following discussion, I assume that the problem of overcompleteness has been properly removed.

Next, mention the relationship between the GCM and the collective motion.
In the matrix representation, the Hill-Wheeler equation (\ref{eq:HW_GCM}) can be written as
\begin{equation}
  \label{eq:HW_mat}
  (H - EI)\bm{f}=0.
\end{equation}
Unless the problem of overcompleteness arises, the square root of the norm kernel $I^{1/2}$ and the inverse $I^{-1/2}$ can be constructed.
Then, the normalized Hill-Wheeler equation is obtained as
\begin{gather}
  \label{eq:HW_col}
  (\tilde{H} - E)\bm{g} = 0, \\
  \tilde{H} = I^{-1/2}HI^{-1/2}, \quad \bm{g}=I^{1/2}\bm{f}.
\end{gather}
This equation looks like a non-local Schr$\ddot{\mathrm{o}}$dinger equation.
A local Schr$\ddot{\mathrm{o}}$dinger-type equation is obtained if one performs a Gaussian overlap approximation (GOA) \cite{Ring_Schuck}, which is a kind of derivative expansion.
From this analogy, it is considered that the GCM can be used for a microscopic description of the collective motion.
In this context, the generator coordinate $q$ is referred to as a collective coordinate.

\subsection{DGCM}
In the GCM, how determining the path is nontrivial, and the empirical method often yields wrong results \cite{PT62}.
This problem is due to the GCM being ill-defined, and some proper restrictions are needed.
The dynamical GCM (DGCM) is an improved version of the GCM by giving it a structure like analytical mechanics \cite{GR80}.
In this section, I review the basic ideas of the DGCM and make some comments on the DGCM.

The DGCM is a kind of GCM that simultaneously considers the conjugate momentum $p$ and the collective coordinate $q$.
Then, a normalized path $|q, p\rangle$ with the two parameters $(q, p)$ is defined as
\begin{equation}
  \label{eq:CC}
  \langle q,p|\overleftarrow{\partial}_q\overrightarrow{\partial}_p - \overleftarrow{\partial}_p\overrightarrow{\partial}_q|q,p\rangle = i,
\end{equation}
where, for example, $\overleftarrow{\partial}_q$ means a partial derivative of $q$ acting on the left side.
This equation is referred to as a conjugation condition, which means something like a canonical commutation relation.
Using the dynamical path, any state contained in the collective subspace of the DGCM is given as in the case of the GCM,
\begin{equation}
  \label{eq:DGCM}
  |\Psi\rangle_{\mathrm{DGCM}} = \iint dqdp\, f(q, p)|q, p\rangle.
\end{equation}
Then, the Hill-Wheeler equation of the DGCM can be derived as
\begin{equation}
  \label{eq:DGCM_HW}
  \iint dqdp\left[
  \langle q',p'|\hat{H}|q,p\rangle - E\langle q',p'|q,p\rangle
  \right]f(q,p)
  = 0.
\end{equation}
Goeke and Reinhard found that the DGCM gives the consistent inertia within the GOA \cite{GR80}.

Consider rewriting the conjugation condition (\ref{eq:CC}) with a Berry curvature.
A Berry connection for the dynamical path is defined as
\begin{equation}
  A := i\langle q,p|d|q,p\rangle
  = i\langle q,p|\frac{\partial}{\partial q}|q,p\rangle dq
  + i\langle q,p|\frac{\partial}{\partial p}|q,p\rangle dp,
\end{equation}
where $d$ is an exterior derivative.
Then, the Berry curvature is derived as
\begin{equation}
\label{eq:BC_original}
  \Omega := dA = i\langle q,p|\overleftarrow{\partial}_q\overrightarrow{\partial}_p - \overleftarrow{\partial}_p\overrightarrow{\partial}_q|q,p\rangle dq\wedge dp.
\end{equation}
One can realize that the conjugation condition means $\Omega$ is a symplectic form in analytical mechanics (standard symplectic form),
\begin{equation}
  \label{eq:BC}
  \Omega = -dq\wedge dp.
\end{equation}
Of course, a similar structure can be derived when $p$ and $q$ are extended to multi-dimensional cases.
Therefore, the DGCM can be understood as the GCM with the standard symplectic structure through the Berry curvature.
One of the remarkable features of the Berry curvature is invariance under the local U(1) gauge transformation $|q, p\rangle \to e^{i\theta(q,p)}|q, p\rangle$.
On the other hand, the collective subspace of the GCM, including the DGCM, is also invariant under it, and in this sense, the GCM has the local U(1) symmetry.
Thus, when considering improvements to the GCM by restricting the path, it is natural to impose conditions to gauge-invariant objects to keep this symmetry, such as the DGCM.
Another well-known gauge-invariant object is a Fubini-Study metric,
\begin{align}
    \label{eq:FSM}
    g_{\mu\nu}(q,p) & := 
    \frac{1}{2}\langle q, p| \overleftarrow{\partial}_{\mu}\overrightarrow{\partial}_{\nu}
    + \overleftarrow{\partial}_{\nu}\overrightarrow{\partial}_{\mu}
    |q, p\rangle
    -
    \langle q, p|\overleftarrow{\partial}_{\mu}|q, p\rangle
    \langle q, p|\overrightarrow{\partial}_{\nu}|q, p\rangle,
\end{align}
where $\mu, \nu = q, p$.
It may be meaningful that the Fubini-Study metric is intimately related to the Berry curvature through a quantum geometric tensor (QGT) \cite{PV80, QGT, GB20, LG21}.

Mention a different representation of the DGCM for a particular case.
In order to determine the path satisfying the condition (\ref{eq:CC}), some boundary condition is necessary.
In nuclear theory, the collective coordinate $q$ is often generated using a constrained variational method \cite{Ring_Schuck}.
That is, the many-body wave function $|q\rangle$ is optimized using the variation under the constraint condition about a Hermitian operator $\hat{Q}$,
\begin{equation}
  \label{eq:const0}
  \langle q|\hat{Q}|q\rangle = q,
\end{equation}
where $|q\rangle$ is assumed to be normalized.
The variation is often performed on some restricted function space, such as the IPA wave function \cite{Ring_Schuck, Bender2003} or the AMD wave function \cite{KK12}.
With this in mind, I introduce the collective coordinate using the constraint condition,
\begin{equation}
  \label{eq:const}
  \langle q|\hat{Q}|q\rangle = q + \mathrm{const.}, \quad q\in \mathcal{M}_C.
\end{equation}
Of course, many states can satisfy this condition, and $|q\rangle$ is not determined uniquely.
However, in the discussion of this subsection, it is not necessary to choose a unique path, and I assume only that $|q\rangle$ is differentiable at every point.
Then, one of the dynamical paths satisfying the conjugation condition (\ref{eq:CC}) is obtained by
\begin{equation}
    \label{eq:DP_simple}
  |q,p\rangle = e^{i\hat{Q}p}|q\rangle, \quad p\in\mathbb{R}.
\end{equation}
It should be noted that even under the current condition, the dynamical path is also not uniquely determined.
For example, given a dynamical path $|q, p\rangle$, the new path $|q,p\rangle_{U} := \hat{U}|q,p\rangle$ with any global unitary operator $\hat{U}$ also satisfies the conjugation condition (\ref{eq:CC}).
Thus, when a transformation keeps the Berry curvature (\ref{eq:BC_original}) invariant, the dynamical path is not uniquely determined in the current approach.
In addition, there may be local unitary transformations that keep the Berry curvature invariant, depending on the properties of $|q, p\rangle$.
In this paper, however, I adopt Eq. (\ref{eq:DP_simple}) for simplicity.
Then, one can realize that the DGCM ansatz (\ref{eq:DGCM}) is equivalent to the generalized double projection ansatz \cite{HH21},
\begin{equation}
  \label{eq:DPF}
  |\Psi\rangle_{\mathrm{DGCM}}=\int_{\mathbb{R}} dq\int_{\mathcal{M}_C} dq' \, \tilde{f}(q;q')|q;q'\rangle,
\end{equation}
where $\tilde{f}(q;q')$ is a new weight function.
This two-parameter state $|q;q'\rangle$ is defined as
\begin{align}
  |q;q'\rangle & =\mathrm{normalization}\times \int_{\mathbb{R}} \frac{dp}{2\pi}e^{i(\hat{Q}-q)p}|q'\rangle \\
  & =: \hat{P}^{\hat{Q}}(q)|q'\rangle,
\end{align}
where $\hat{P}^{\hat{Q}}(q)$ is the projection operator about $\hat{Q}$ such that
\begin{equation}
  \hat{Q}\hat{P}^{\hat{Q}}(q)|q'\rangle = q\hat{P}^{\hat{Q}}(q)|q'\rangle.
\end{equation}
Thus, $q$ of $\hat{P}^{\hat{Q}}(q)$ means an eigenvalue of the Hermitian operator $\hat{Q}$.
This representation is an extension of the double projection method \cite{PT62}.
Note that this scheme leaves ambiguity in determining the state $|q\rangle$ that satisfies the constraint condition (\ref{eq:const}).
It can be easily verified that the Fubini-Study metric (\ref{eq:FSM}) is $p\,$-independent for the dynamical path (\ref{eq:DP_simple}).

\section{Validity of the DGCM}
\subsection{Preferred structure}
The essential degree of freedom in the DGCM (or the GCM) is not the path but the collective subspace.
Thus, the validity of the DGCM (or the GCM) should be discussed with the collective subspace.
However, that is generally difficult since the collective motion is not strictly defined. When considering such a complex system, it seems to be a fundamental idea in physics to consider a simple system as a starting point.
Therefore, taking the simplest case of the collective motion, I define the preferred structure for describing it.

Consider the quantum mechanics of a many-particle system.
Due to some continuous symmetries of the system, there may be a specific canonical transformation that separates the degrees of freedom.
For example, the entire nucleus's translational motion and rotational motion correspond to this case.
Therefore, consider a separable (non-interacting) Hamiltonian for the collective motion,
\begin{equation}
  \label{eq:NIH}
  \hat{H}_{\mathrm{tot}}=\hat{H}_C\otimes \hat{\bm{1}}_{NC}
  +\hat{\bm{1}}_{C}\otimes \hat{H}_{NC},
\end{equation}
where $\hat{H}_{C(NC)}$ and $\hat{\bm{1}}_{C(NC)}$ represent the collective (non-collective) part of the Hamiltonian and the identical operator, respectively.
The non-collective degrees of freedom can be read as internal ones.
Then, the total Hilbert space $\mathcal{H}_{\mathrm{tot}}$ is assumed to be decomposed with the tensor product as
\begin{equation}
  \mathcal{H}_{\mathrm{tot}}=\mathcal{H}_C \otimes \mathcal{H}_{NC}.
\end{equation}
Of course, there can be some symmetry of the particles concerning interchangeability, and it may not allow for the simple tensor product.
However, I assume this form to deal with simple cases, such as the translational motion and the rotational motion.
Since the Hamiltonian (\ref{eq:NIH}) has no interaction between the subsystems $C$ and $NC$, its eigenvalue problem can be solved formally as
\begin{gather}
  \label{eq:SES}
  |n, \mu\rangle = |n\rangle_{C}\otimes |\mu\rangle_{NC}, \\
  \label{eq:SE}
  \hat{H}_{\mathrm{tot}}|n, \mu\rangle
  = (E^n_{C} + E^{\mu}_{NC})|n, \mu\rangle
  =: E_{\mathrm{tot}}^{n, \mu}|n, \mu\rangle,
\end{gather}
where 
\begin{gather}
  \hat{H}_C|n\rangle_{C} = E_{C}^{n}|n\rangle_{C}, \quad |n\rangle_{C} \in \mathcal{H}_C, \\
  \hat{H}_{NC}|\mu\rangle_{NC} = E_{NC}^{\mu}|\mu\rangle_{NC}, \quad |\mu\rangle_{NC} \in \mathcal{H}_C.
\end{gather}
The remarkable feature of this solution is that all eigenstates are {\it separable}.
If one use a separable collective subspace,
\begin{equation}
  \label{eq:SCS}
  \mathcal{H}_{\mathrm{DGCM}} = \mathcal{H}^{\mathrm{sub}}_C \otimes \mathcal{H}^{\mathrm{sub}}_{NC}
  , \quad \mathcal{H}^{\mathrm{sub}}_C \subseteq \mathcal{H}_C
  , \quad \mathcal{H}^{\mathrm{sub}}_{NC} \subseteq \mathcal{H}_{NC},
\end{equation}
all solutions to the Hill-Wheeler equation with the Hamiltonian (\ref{eq:NIH}) are separable states.
Then, I define Eq. (\ref{eq:SCS}) as the preferred structure of the collective subspace in terms of keeping the solutions separable and call it a {\it separable} structure in this paper.
Therefore, this paper aims to show that the DGCM contains a framework that yields the separable collective subspace.

The separable collective subspace can also be applied to the systems where the adiabatic approximation works well.
Considering a system with the interaction between the subsystems $C$ and $NC$, the eigenstates have quantum entanglement.
However, due to the scale difference between two subsystems, the eigenstates of the total system may be well approximated by separable states.
When investigating such systems with the empirical GCM, in which the structure of the collective subspace is often unknown, the risk of artificial quantum entanglement arises.
To avoid this problem and treat the adiabatic systems properly, the separable structure (\ref{eq:SCS}) is significant.

Finally, mention a critical assumption in the DGCM (or GCM).
It is not easy to find a canonical transformation leading to Eq. (\ref{eq:NIH}) except for trivial cases, such as the translational motion.
The existence of a collective operator $\hat{Q}_C$, which is relevant to the collective degree of freedom and should be a Hermitian operator on $\mathcal{H}_C$ with no degeneracy,  is assumed in this paper.
The SCC can derive such a collective operator from a given Hamiltonian \cite{MM80, M84, MN00, NM06}.
However, note that it is conceptually different from the GCM (DGCM), and the obtained operator depends on the collective coordinate in general.
Thus, many current GCM calculations have two inherent problems: whether the empirically selected degrees of freedom are collective actually and whether the generated collective subspace is appropriate.
These two problems are intricately intertwined and complicate the argument for the collective motion.
Therefore, this paper's motivation is to establish a basic framework for determining whether the assumed operator $\hat{Q}_C$ is relevant to the collective motion by appropriately restricting the structure of the collective subspace.

\subsection{Locally separable structure}
To discuss the validity of the DGCM,
I adopt the double projection form of the DGCM (\ref{eq:DPF}) with the operator $\hat{Q} = \hat{Q}_C\otimes \hat{\bm{1}}_{NC}$.
Then the projected state $|q;q'\rangle$ can be decomposed into
\begin{equation}
  \label{eq:DCP_ansatz}
  |q; q'\rangle = |q\rangle_{C}\otimes |q;q'\rangle_{NC}
  ,  \quad |q\rangle_{C}\in \mathcal{H}_C, \quad
  |q;q'\rangle_{NC}\in \mathcal{H}_{NC},
\end{equation}
where $\{|q\rangle_C\}_q$ are the normalized eigenstates of $\hat{Q}_C$.
Note that $\overline{\mathrm{Span}}(\{|q\rangle_C\}_q) = \mathcal{H}_C$ because the eigenstates of $\hat{Q}_C$ form a complete set in $\mathcal{H}_C$.
Under the above decomposition, consider the collective subspace in the DGCM,
\begin{equation}
  \mathcal{H}_{\mathrm{DGCM}} := \overline{\mathrm{Span}}(\{|q;q'\rangle \,|\,q\in \mathcal{M}_C, \, q'\in\mathbb{R}\}).
\end{equation}
The structure is specified using a direct sum and a tensor product as
\begin{equation}
  \label{eq:TPS}
  \mathcal{H}_{\mathrm{DGCM}} =
  \bigoplus_{q\in \mathcal{M}_{C}}\mathcal{H}_C^q\otimes \mathcal{H}_{NC}^q,
\end{equation}
where
\begin{gather}
  \mathcal{H}^{q}_{C}
  := \mathrm{span}(\{|q\rangle_C \}) \subset \mathcal{H}_{C}, \\
  \mathcal{H}^{q}_{NC} := \overline{\mathrm{Span}}(\{|q;q'\rangle_{NC}\,|\,
  q' \in \mathbb{R}
  \}) \subseteq \mathcal{H}_{NC}.
\end{gather}
Therefore, $\mathcal{H}_{\mathrm{DGCM}}$ does not always have the separable structure (\ref{eq:SCS}).
However, it can be considered separable only for a sector within the neighborhood of a particular $q$, and in this sense, it is {\it locally} separable.
Thus, it is expected to give a good approximation for localized solutions in the coordinate $q$.
If $\mathcal{H}_{NC}^q$ is independent of $q$, it becomes {\it globally} separable.
It should be emphasized that the collective subspace can be limited to this form from only the current setting.
At the moment, even a function space used to draw the path $|q\rangle$ is not specified.
In the GCM case, however, the structure strongly depends on the properties of the function space, and one can not guarantee even the locally separable structure (\ref{eq:TPS}) in general.
If one tries to obtain the locally separable structure, it is necessary to consider the momentum introduced in the form of Eq. (\ref{eq:DP_simple}).
One of the advantages of the DGCM is that the conjugation condition includes the conjugate momentum naturally.
Note that the coordinate $q$, introduced as the collective coordinate, seems to behave like a non-collective degree of freedom.

\subsection{Globally separable structure}
This subsection shows that adding several conditions yields the separable collective subspace (\ref{eq:SCS}).
To determine $|q\rangle$ uniquely, consider the momentum operator $\hat{P} = \hat{P}_C \otimes \hat{\bm{1}}_{NC}$ conjugate to $\hat{Q}$.
Then, the path $|q\rangle$ can be generated as
\begin{equation}
\label{eq:ME}
  |q\rangle = e^{-i\hat{P}q}|\psi\rangle, \quad
   |\psi\rangle \in \mathcal{H}_{\mathrm{tot}},
\end{equation}
where $|\psi\rangle$ is some reference state.
Of course, it can be easily seen that the path $|q\rangle$ satisfies the constrained condition (\ref{eq:const}).
The decomposition of the reference state is written as
\begin{equation}
\label{eq:DCP}
  |\psi\rangle = \int_{\mathbb{R}} dq\,|q\rangle_C\otimes |q;\psi\rangle_{NC},
  \quad
   |q;\psi\rangle_{NC} \in \mathcal{H}_{NC}.
\end{equation}
Note that the non-collective states can include the case $|q;\psi\rangle_{NC} = 0$ for some $q$.
The canonical commutation relation $[\hat{Q}, \hat{P}]=i\hat{\bm{1}}_C\otimes \hat{\bm{1}}_{NC}$ yields the decomposition of the path,
\begin{equation}
  |q\rangle = \int _{\mathbb{R}} dq'\, |q' + q\rangle_C \otimes |q';\psi\rangle_{NC}.
\end{equation}
Then, performing the projection of $\hat{Q}$ derives the projected state,
\begin{equation}
  \label{eq:TB}
  |q;q'\rangle = |q\rangle_C\otimes |q-q' ; \psi\rangle_{NC}
\end{equation}
Therefore, the double projection form of the DGCM ansatz can be evaluated as
\begin{equation}
  \label{eq:SGA}
  |\Psi \rangle_{\mathrm{DGCM}} = \int_{\mathbb{R}} dq\int _{\mathcal{M}_C} dq'\,\tilde{f}(q;q')|q\rangle_C\otimes |q' ; \psi\rangle_{NC},
\end{equation}
where I assume that the interval of the integral for the non-collective degrees of freedom is invariant under the change of the variable $q - q' \to q'$.
This assumption implies the manifold $\mathcal{M}_C$ is $\mathbb{R}$ or $S^1$ in a connected one-dimensional manifold.
Equation (\ref{eq:SGA}) shows that the non-collective state $|q';\psi\rangle_{NC}$ is no longer dependent on the collective part, and this result identifies the structure of the collective subspace,
\begin{equation}
  \label{eq:SHS}
  \mathcal{H}_{\mathrm{DGCM}} = \mathcal{H}_{C} \otimes \mathcal{H}^{\psi}_{NC}, \quad
  \mathcal{H}^{\psi}_{NC} 
  := \overline{\mathrm{Span}}(\{|q'; \psi\rangle_{NC}\, |\, q' \in \mathcal{M}_C\}).
\end{equation}
Therefore, based on the DGCM, it is possible to systematically construct the collective subspace with the separable structure (\ref{eq:SCS}).
Notably, this argument does not rely on the Hamiltonian.
Therefore, if one has information on both $\hat{Q}$ and $\hat{P}$ for a given collective motion, the separable collective subspace can be obtained systematically.
Then, the Fubini-Study metric for the dynamical path is evaluated as
\begin{align}
    g_{\mu\nu} & = 
    \begin{pmatrix}
    \langle(\hat{P} - \langle \hat{P}\rangle)^2\rangle & 
    -\frac{1}{2}\langle\{\hat{Q} - \langle\hat{Q}\rangle , \hat{P} - \langle \hat{P}\rangle \}\rangle \\
    -\frac{1}{2}\langle\{\hat{Q} - \langle\hat{Q}\rangle , \hat{P} - \langle \hat{P}\rangle \}\rangle 
    & \langle(\hat{Q} - \langle \hat{Q}\rangle)^2\rangle
    \end{pmatrix}
    _{\mu\nu},
\end{align}
where the expectation value $\langle \psi|\cdot|\psi\rangle$ is abbreviated as $\langle \,\cdot\, \rangle $, and it depends only on the reference state, not on the coordinate $(q,p)$.

This method has the ambiguity of the reference state $|\psi\rangle$.
It should be determined appropriately according to the energy scale of interest.
In the low-energy dynamics, one can prepare it such that $\langle \psi|\hat{H}|\psi\rangle$ is small, for example, using a variational method.
Of course, this method can be extended to the case of multi-reference states $\{|\psi_1\rangle, |\psi_2\rangle, \cdots\}$.
Then, the collective subspace can be evaluated as $\mathcal{H}_{C} \otimes \left(\mathcal{H}^{\psi_1}_{NC} + \mathcal{H}^{\psi_2}_{NC} + \cdots \right)$, and the non-collective part of the collective subspace becomes larger.

The above method has the problem that the specific form of the momentum operator $\hat{P}$ is not always identified, and this method by itself is not practical.
However, the variational method enables the obtainment of the separable collective subspace (\ref{eq:SHS}) without specific information on $\hat{P}$ under several conditions.
A variational equation for the energy expectation value is written as
\begin{equation}
  \label{eq:VE}
  \delta_{\mathcal{F}} \langle \psi|\hat{H}|\psi\rangle = 0,
\end{equation}
where the variation $\delta_{\mathcal{F}}$ is performed on a function space $\mathcal{F}$, in which any function is normalized.
For example, $\mathcal{F}$ is the set of the normalized IPA wave functions or the normalized AMD wave functions.
The solution to Eq. (\ref{eq:VE}) and the expectation value of $\hat{Q}$ are denoted by $|\psi_0\rangle$ and $q_0 = \langle \psi_0| \hat{Q}|\psi_0\rangle$, respectively.
It should be noted that the solution is not always uniquely determined.
This problem arises when a unitary transformation keeps the Hamiltonian and the variational space invariant.
One can resolve it to restrict the variational space, for example, by inserting a constraint on some operator that is not invariant to the unitary transformation.
Thus, in the following discussion, $\mathcal{F}$ is the restricted variational space where the solution to Eq. (\ref{eq:VE}) is uniquely determined.
Consider imposing a constraint condition on this variational equation,
\begin{equation}
  \label{eq:CVE1}
  \delta_{\mathcal{F}} \langle \psi|\hat{H} - \lambda(\hat{Q} - q)|\psi\rangle = 0,\quad q\in\mathbb{R},
\end{equation}
where $\lambda$ is a Lagrange multiplier.
The solution is denoted by $|q\rangle$ and, of course, satisfies $\langle q|\hat{Q}| q\rangle = q$.
For the function space $\mathcal{F}$, I then assume
\begin{equation}
  \label{eq:ASM1}
  ^{\forall}|\psi\rangle \in \mathcal{F}
  \quad \Rightarrow \quad
  e^{-i\hat{P}a}|\psi\rangle \in \mathcal{F}
  \quad
  \mathrm{for \ all \ } a\in \mathbb{R}.
\end{equation}
This assumption leads to the result that Eq. (\ref{eq:CVE1}) is equivalent to
\begin{equation}
  \label{eq:CVE2}
  \delta_{\mathcal{F}} \langle \psi|e^{i\hat{P}a}\{\hat{H} - \lambda(\hat{Q} - q)\}e^{-i\hat{P}a}|\psi\rangle = 0
  \quad
  \mathrm{for \ all \ } a \in \mathbb{R}.
\end{equation}
Considering the separable Hamiltonian (\ref{eq:NIH}), I further assume that the collective part  $\hat{H}_C$ is quadratic for $\hat{Q}$,
\begin{equation}
  \label{eq:ASM2}
  \hat{H}_C\otimes \hat{\bm{1}}_{NC} = \alpha\hat{Q}^2 + V(\hat{P}),
\end{equation}
where $\alpha$ is a non-zero real number, and $V(\cdot)$ is an arbitrary function.
Of course, the first-order term of $\hat{Q}$ can be included, but it can be eliminated by completing the square and need not be considered explicitly.
Then, evaluating Eq. (\ref{eq:CVE2}) yields
\begin{equation}
  \delta_{\mathcal{F}} \langle \psi|\hat{H} +2a\alpha \hat{Q} + \alpha a^2   - \lambda(\hat{Q} + a - q)|\psi\rangle = 0.
\end{equation}
From the variation for $\lambda$, the constraint condition $\langle \psi|\hat{Q}|\psi\rangle = q - a$ is derived, and substituting it leads to
\begin{equation}
  \langle \psi|\hat{H} +2a\alpha \hat{Q} + \alpha a^2   - \lambda(\hat{Q} + a - q)|\psi\rangle
  =
  \langle \psi|\hat{H}|\psi\rangle
  + a\alpha(2q - a).
\end{equation}
Then, $\langle \psi|\hat{H}|\psi\rangle \geq \langle \psi_0|\hat{H}|\psi_0\rangle$ for any $|\psi \rangle \in \mathcal{F}$
, and the equal sign is valid only when $|\psi\rangle = |\psi_0\rangle$.
If one takes $a = q - q_0$, Eq. (\ref{eq:CVE2}) yields $|\psi\rangle = |\psi_0\rangle$, and hence the solution to Eq. (\ref{eq:CVE1}) is $|\Psi\rangle = |q - q_0\rangle = e^{-i\hat{P}(q-q_0)}|\psi_0\rangle$.
Therefore, under the two assumptions (\ref{eq:ASM1}, \ref{eq:ASM2}), the path $|q\rangle$ defined by Eq. (\ref{eq:ME}) can be obtained without specific information of $\hat{P}$.
One can regard this result as an extension of the discussion in the double projection \cite{PT62}, which is based on symmetries.
It is trivial that this approach works well even when $\hat{H}_C\otimes \hat{\bm{1}}_{NC} = \beta\hat{Q} + \gamma + V(\hat{P}), \quad \beta, \gamma \in\mathbb{R}$.

This approach is superior because it is feasible to perform numerical calculations.
For example, several groups have performed it as the double projection method for the rotational motion \cite{BR15,EB16, ST15, Shimada2016, Ushitani2019}, and the results reproduce experimental values well.
In the study for the quadrupole oscillation based on this approach \cite{HH22},  the sum rule result suggests the separable structure of the collective subspace.
In addition, one can extract information on $\hat{P}$ from the numerical results if Eqs. (\ref{eq:NIH}, \ref{eq:ASM2}) are satisfied approximately well for a small $|q - q_0|$.
Then, performing Eq. (\ref{eq:ME}) with this $\hat{P}$ yields the separable collective subspace that does not depend on the structure of the total Hamiltonian.

Note that one can make a similar discussion even if $\hat{P}$ and $\hat{Q}$ are interchangeable.
Then, the assumption (\ref{eq:ASM2}) becomes $\hat{H}_C\otimes \hat{\bm{1}}_{NC} = \alpha\hat{P}^2 + V(\hat{Q})$, which includes an anharmonic potential.
In quantum mechanics, our interest is often limited to the second-order of the momentum.
Thus, if the momentum operator $\hat{P}$ is specified in advance, this method can be applied to a broader class of realistic collective motions beyond vibration.

\subsection{Symplectic transformation}
In the SCC, the collective operators $(\hat{Q}, \hat{P})$ depend on the canonical coordinates $(q, p)$.
However, the local collective operators are not always relevant for the collective subspace.
Consider a linear transformation for the collective operators,
\begin{equation}
    \hat{\bm{A}} \, \to \, \hat{\bm{A}}' = S\hat{\bm{A}}
    ,  \quad \hat{\bm{A}} := \begin{pmatrix}
    \hat{Q} \\
    \hat{P}
    \end{pmatrix}.
\end{equation}
Then, the condition that the new collective operators $\hat{\bm{A}}'$ are also Hermitian and satisfy the canonical commutation relation leads to $S$ being a real symplectic matrix; namely, $S \in \mathrm{Sp}(2, \mathbb{R})$.
In the two-dimensional case, the symplectic group $\mathrm{Sp}(2, \mathbb{R})$ is equal to a special linear group $\mathrm{SL}(2, \mathbb{R})$.
The dynamical path given by Eq. (\ref{eq:ME}) is written as
\begin{equation}
    \label{eq:DPA}
    |q,p\rangle =e^{-i\bm{a}^T\sigma \hat{\bm{A}}}|\psi\rangle,
    \quad
    \bm{a} = \begin{pmatrix}
    q \\
    p
    \end{pmatrix},
    \quad
    \sigma =
    \begin{pmatrix}
    0 & 1 \\
    -1 & 0
    \end{pmatrix},
\end{equation}
where the local phase degree of freedom is irrelevant for the DGCM and omitted.
In this subsection, I assume $\mathcal{M}_C=\mathbb{R}$ for simplicity, namely $\bm{a} \in \mathbb{R}^2$.
It is trivial that the global symplectic transformation 
$e^{-i\bm{a}^T\sigma \hat{\bm{A}}}|\psi\rangle \, \to \, e^{-i\bm{a}^T\sigma \hat{\bm{A}}'}|\psi\rangle$ does not change the collective subspace;
\begin{equation}
    \overline{\mathrm{Span}}(\{e^{-i\bm{a}^T\sigma \hat{\bm{A}}}|\psi\rangle \, |\,
    \bm{a}\in \mathbb{R}^2\})
    =
    \overline{\mathrm{Span}}(\{e^{-i\bm{a}^T\sigma \hat{\bm{A}}'}|\psi\rangle \, |\,
    \bm{a}\in \mathbb{R}^2\}).
\end{equation}
This transformation is just a redefinition of the coordinate-independent canonical operator pair for $\mathcal{H}_C$.
Of course, $S$ also includes the operation of swapping $\hat{Q}$ and $\hat{P}$.

Next, extend the global symplectic transformation to a local case,
\begin{equation}
\label{eq:LST}
    \hat{\bm{A}} \, \to \, \hat{\bm{A}}'(\bm{a}) = S(\bm{a})\hat{\bm{A}},
    \quad
    S(\bm{a}) \in \mathrm{Sp}(2, \mathbb{R})
    \quad 
    \mathrm{for \ all \ }
    \bm{a} \in \mathbb{R}^2.
\end{equation}
The question is how this transformation changes the collective subspace, 
$\mathcal{H}_{\mathrm{DGCM}}[S] := \overline{\mathrm{Span}}(\{e^{-i\bm{a}^T\sigma \hat{\bm{A}}'(\bm{a})}|\psi\rangle \, |\,\bm{a}\in \mathbb{R}^2\})$
To see this, consider the corresponding transformation for the DGCM ansatz,
\begin{equation}
    \label{eq:TDA}
    |\Psi\rangle_{\mathrm{DGCM}}
    \ \to \ 
    |\Psi[S]\rangle_{\mathrm{DGCM}}
    =
    \int_{\mathbb{R}^2} d^2a\,f(\bm{a}) e^{-i\bm{a}^T\sigma \hat{\bm{A}}'(\bm{a})}|\psi\rangle.
\end{equation}
The condition $S \in \mathrm{Sp}(2, \mathbb{R})$ equals $S^T(\bm{a})\sigma S(\bm{a}) = \sigma$, and that yields
\begin{equation}
    \label{eq:BST}
    \bm{a}^T \sigma \hat{\bm{A}}
    \  \to \ 
    \bm{a}^T \sigma \hat{\bm{A}}'(\bm{a})
    = (S^{-1} \bm{a})^T \sigma \hat{\bm{A}}
    =:\bm{b}^T \sigma \hat{\bm{A}}.
\end{equation}
Thus, the change of the variable $\bm{a} \to \bm{b}$ leads to
\begin{equation}
    \label{eq:COV}
    |\Psi[S]\rangle_{\mathrm{DGCM}}
    =
    \int_{\mathbb{R}^2}d^2 b \,
    |\det J(\bm{a}(\bm{b}))|^{-1}
    \,f(\bm{a}(\bm{b}))
    e^{-i{\bm{b}}^T\sigma \hat{\bm{A}}}|\psi\rangle,
\end{equation}
where the Jacobian matrix $J$ is defined as
\begin{equation}
    J_{\mu\nu}
    =
    \frac{\partial b_{\nu}}{\partial a_{\mu}}
    =
    \sum_{\lambda}
    \frac{\partial S^{-1}_{\nu\lambda}}{\partial a_{\mu}}a_{\lambda}
    +
    S^{-1}_{\nu\mu}
    .
\end{equation}
Of course, the transformation is possible only when $J$ is a non-singular matrix.
However, as seen later, this condition is guaranteed from the conjugation condition, so I proceed with the discussion, assuming that now.
In addition, I assume the image of the function $\bm{b}(\cdot)$ is $\mathbb{R}^2$ and that it is injective.
Since the weight function $f$ is arbitrary, $g(\bm{b}) = |\det J|^{-1}\,f$ can also be taken arbitrarily.
Then, the transformed DGCM ansatz (\ref{eq:TDA}) is written as
\begin{equation}
    |\Psi[S]\rangle_{\mathrm{DGCM}}
    =
    \int_{\mathbb{R}^2}d^2 b
    \,g(\bm{b})
    e^{-i{\bm{b}}^T\sigma \hat{\bm{A}}}|\psi\rangle,
\end{equation}
and the transformed collective subspace is evaluated as
\begin{equation}
    \mathcal{H}_{\mathrm{DGCM}}[S]
    = \mathcal{H}_{C}\otimes \mathcal{H}_{NC}^{\psi}.
\end{equation}
Therefore, the globally separable collective subspace (\ref{eq:SHS}) is invariant under the local symplectic transformation (\ref{eq:LST}) under the several conditions.

When considering the symplectic transformation within the context of the DGCM, one must impose the conjugation condition (\ref{eq:CC}) for the transformed dynamical path.
It is useful that Eq. (\ref{eq:BST}) yields
\begin{equation}
    |q,p\rangle_{S} := e^{-i\bm{b}^T \sigma \hat{\bm{A}}}|\psi\rangle
    = |b_1, b_2\rangle.
\end{equation}
Namely, the transformed path $|q,p\rangle_S$ and the original path $|q,p\rangle$ are represented using the same function form $|\,\cdot\,, \,\cdot\,\rangle$.
Then, the Berry curvature is transformed as
\begin{equation}
    B(\bm{a}) = -da_{1} \wedge da_{2}
    \ \to \ 
    B(\bm{b}) = -db_{1} \wedge db_{2}.
\end{equation}
Therefore, the conjugation condition (\ref{eq:CC}) imposes that the transformation $\bm{a} \to \bm{b}$ is a canonical transformation,
\begin{equation}
    \label{eq:CT}
    da_{1} \wedge da_{2} = db_{1} \wedge db_{2}
    =
    \det J(\bm{a}) \, da_1 \wedge da_2.
\end{equation}
One can realize that the conjugation condition guarantees $|\det J| = 1$, and the change of the variable (\ref{eq:COV}) is possible.
Using an analytical mechanics technique, the solution to this equation is obtained with a generating function $W(a_1, b_1)$;
\begin{equation}
    a_2 = \frac{\partial W (a_1, b_1)}{\partial a_1}
    ,\quad 
    b_2 = -\frac{\partial W (a_1, b_1)}{\partial b_1}.
\end{equation}
Therefore, the equations that $S$ must satisfy can be summarized as
\begin{equation}
\label{eq:STCC}
\left\{ \,
    \begin{aligned}
    & a_2 = \frac{\partial W (a_1, b_1)}{\partial a_1} \biggr|_{b_1 = S_{22}a_1 - S_{12}a_2}
    \\
    & S_{11}a_2 - S_{21}a_1 = 
    -\frac{\partial W (a_1, b_1)}{\partial b_1} \biggr|_{b_1 = S_{22}a_1 - S_{12}a_2}
    \\
    & \det S = 1
    \end{aligned}
\right.,
\end{equation}
where $S^{-1} = -\sigma S^{T}\sigma$ and $S\in \mathrm{Sp}(2, \mathbb{R}) \Leftrightarrow \det S = 1$ are used.
Since there are only three conditions for five unknown functions $(S_{11}, S_{12}, S_{21}, S_{22}, W)$, Eq. (\ref{eq:STCC}) has an infinite number of solutions.
If one considers an extension of the current $2$-dimensional discussion to a $2n$-dimensional case, there are $n(2n + 1)$ independent components for $\mathrm{Sp}(2n, \mathbb{R})$
, and $2n-1$ conditions are imposed from the conjugation condition.
There are $2n^2 - n + 1$ remaining unknown components,
and hence the DGCM has this much redundancy when using the canonical operators.

\subsection{VAP}
In variational calculations, such as the mean-field theory, a property of a function space $\mathcal{F}$ often brokes the system's symmetries.
In order to restore these, a variation after projection method (VAP) is often used \cite{Ring_Schuck}.
As the name implies, the original trial function is projected onto an eigenstate of the symmetry operator in advance and optimized with the variation.
Typically, the VAP is performed for the particle number operator \cite{Sheikh2002}, the angular momentum operator \cite{K98}, and the translational momentum operator \cite{RS04}.

The analytical technique in the previous subsection can be applied to the VAP.
Consider the VAP function space about the Hermitian operator $\hat{Q} = \hat{Q}_C\otimes \hat{\bm{1}}_{NC}$,
\begin{equation}
   \mathcal{F}_{\mathrm{VAP}} :=\{\hat{P}^{\hat{Q}}(q)|\psi\rangle\,|\,|\psi\rangle  \in \mathcal{F}, \,q\in \mathbb{R}\},
\end{equation}
where $\mathcal{F}$ is an original function space.
Note that the projection operator is not injective, and duplication can occur.
Furthermore, it is possible that $\hat{P}^{\hat{Q}}(q)|\psi\rangle = 0$.
However, such duplicate elements and zeroes are not significant in the structure of the VAP function space and are ignored in the following discussion.
Practically, the variation is performed only in the restricted sector with the eigenvalue $q$,
\begin{gather}
    \mathcal{F}_{\mathrm{VAP}}(q) := \{\hat{P}^{\hat{Q}}(q)|\psi\rangle\,|\,|\psi\rangle  \in \mathcal{F}\}, \\
    \mathcal{F}_{\mathrm{VAP}} = \bigcup_{q\in\mathbb{R}}\mathcal{F}_{\mathrm{VAP}}(q).
\end{gather}
Then, the variational equation is formulated as
\begin{equation}
    \delta_{\mathcal{F}_{\mathrm{VAP}}(q)}\langle \Psi|\hat{H}|\Psi\rangle
    =
    \delta_{\mathcal{F}}\langle \psi|\hat{P}^{\hat{Q}}(q)\hat{H}\hat{P}^{\hat{Q}}(q)|\psi\rangle = 0,
\end{equation}
where $|\Psi\rangle \in \mathcal{F}_{\mathrm{VAP}}(q)$ in this notation.
Using the decomposition (\ref{eq:DCP}), $\mathcal{F}_{\mathrm{VAP}}(q)$ is evaluated as
\begin{equation}
    \mathcal{F}_{\mathrm{VAP}}(q)
    =
    \{|q\rangle_{C}\otimes |q;\psi\rangle_{NC}\,|\,
    |q;\psi\rangle_{NC} \in \mathcal{F}_{NC}(q)
    \},
\end{equation}
where the function space $\mathcal{F}_{NC}(q) \subseteq \mathcal{H}_{NC}$ is the non-collective part of $\mathcal{F}_{\mathrm{VAP}}(q)$.
In general,  $\mathcal{F}_{\mathrm{VAP}}(q)$ depends on $q$, and hence  $\mathcal{F}_{\mathrm{VAP}}$ has a {\it twisted} structure.
I assume Eq. (\ref{eq:ASM1}) and obtain,
\begin{equation}
  |\phi\rangle :=e^{-i\hat{P}q}|\psi\rangle = \int_{\mathbb{R}} dq'\,|q'+q\rangle_C \otimes |q'; \psi\rangle_{NC}
  \in \mathcal{F}.
\end{equation}
This equation shows that for any $|\psi\rangle \in \mathcal{F}$, $q\in\mathbb{R}$, and $q' \in\mathbb{R}$
, there exists $|\phi\rangle \in \mathcal{F}$ such that
\begin{equation}
  |q'; \psi\rangle_{NC}
  =
  |q'+q; \phi\rangle_{NC}.
\end{equation}
This proposition immediately reveals 
\begin{equation}
    \mathcal{F}_{NC}(q) = \mathcal{F}_{NC}(q') =:\mathcal{F}_{NC}
    \quad
    \mathrm{for \ all \ }q, q' \in\mathbb{R}.
\end{equation}
Therefore, the non-collective part of the VAP function space does not depend on $q$, and $\mathcal{F}_{\mathrm{VAP}}$ is written as
\begin{equation}
\label{eq:UTV}
    \mathcal{F}_{\mathrm{VAP}}
    =
    \{
    |q\rangle_C \otimes | \psi\rangle_{NC} \, |\,
    |q\rangle_C \in \mathcal{F}_{\mathrm{ES}}(\hat{Q}_C), \, 
    | \psi\rangle_{NC} \in \mathcal{F}_{NC}
    \},
\end{equation}
where $\mathcal{F}_{\mathrm{ES}}(\hat{Q}_C) \subset \mathcal{H}_C$ is the set of the eigenstates of $\hat{Q}_C$.
Then, {\it untwisted} $\mathcal{F}_{\mathrm{VAP}}$ can be regarded as 
$\mathcal{F}_{\mathrm{ES}}(\hat{Q}_C) \times \mathcal{F}_{NC}$ and, in this sense, a product space {\it globally}.

Although deviating from the DGCM framework, the untwisted VAP is useful for preparing the separable collective subspace (\ref{eq:SCS}).
In this paper, I discuss the case of using the constrained variational calculation.
Consider the separable Hamiltonian (\ref{eq:NIH}) and a non-collective Hermitian operator $\hat{X}:=\hat{\bm{1}}_C\otimes \hat{X}_{NC}$.
Then, the constrained variational method for $\hat{X}$ with the untwisted VAP function space is defined as
\begin{equation}
    \delta_{\mathcal{F}_{\mathrm{VAP}}(q)}\langle \Psi|\hat{H} - \lambda (\hat{X} - x)|\Psi\rangle
    = 0,
\end{equation}
where the solution is denoted by $|q\rangle_C\otimes |q;x\rangle_{NC}$.
It is trivial that $|q;x\rangle_{NC}$ does not depend on $q$.
Thus, the collective subspace with the untwisted VAP + constrained variational method has the separable structure.
One can realize that this result includes a generalization of the argument in restoring the symmetries with the VAP \cite{Ring_Schuck}.

\subsection{Entanglement entropy}
The above discussions are based on the existence of the collective operator $\hat{Q}$ that leads to the separable Hamiltonian (\ref{eq:NIH}).
However, except in cases involving the symmetries, it is not easy to know $\hat{Q}$ in advance, and an empirically chosen $\hat{Q}$ is used in actual calculations.
Therefore, it is necessary to ascertain whether the assumed $\hat{Q}$ is collective in the sense of Eq. (\ref{eq:NIH}) or the adiabatic approximation.
In this subsection, I introduce and formulate a convenient tool based on the DGCM with the separable collective subspace.

Consider evaluating the degree of entanglement for a solution to the Hill-Wheeler equation $|\Psi\rangle \in \mathcal{H}_C\otimes \mathcal{H}^{\mathrm{sub}}_{NC}$.
Then the entanglement entropy is useful \cite{EC10}, which is defined as
\begin{equation}
  S[\Psi] :=- \Tr_C\hat{\rho}_C\log\hat{\rho}_C, \quad
  \hat{\rho}_C := \Tr_{NC}|\Psi\rangle\langle\Psi|,
\end{equation}
where $|\Psi\rangle$ is a DGCM solution, and $\Tr_{C}$ and $\Tr_{NC}$ mean partial traces over $\mathcal{H}_C$ and $\mathcal{H}^{\mathrm{sub}}_{NC}$, respectively.
If the motion associated with $\hat{Q}$ has the separable Hamiltonian (\ref{eq:NIH}), $|\Psi\rangle$ is the separable state, which leads to $S[\Psi] = 0$.
On the other hand, if there is a non-negligible interaction in the Hamiltonian, then $S[\Psi] \neq 0$.
When performing the partial traces, a complete orthonormal set is useful.
Consider the basis representation of the density operator,
\begin{equation}
  R_{\mu\nu}^{nm} :=\langle n, \mu|\Psi\rangle\langle\Psi|m, \nu\rangle, \quad
  |n, \mu\rangle :=|n\rangle_C\otimes |\mu\rangle^{\mathrm{sub}}_{NC},
\end{equation}
where $|n\rangle_C$ and $|\mu\rangle^{\mathrm{sub}}_{NC}$ are complete orthonormal sets of $\mathcal{H}_{C}$ and $\mathcal{H}^{\mathrm{sub}}_{NC}$, respectively.
Then, the reduced density matrix can be derived as
\begin{equation}
  \langle n|\hat{\rho}_C|m\rangle = \sum_{\mu}R_{\mu\mu}^{mn} :=R_C^{mn}.
\end{equation}
Since $R_C^{mn}$ is a semi-positive definite Hermitian matrix, it can be diagonalized as $R_C = U\mathrm{diag}\{\lambda_1, \lambda_2, \cdots\}U^{\dagger}, \, {}^{\forall}\lambda_m \geq 0$ with the unitary matrix $U$.
Therefore, the entanglement entropy is evaluated as
\begin{equation}
  S[\Psi] = -\sum_{m}\lambda_m\log\lambda_m.
\end{equation}

The problem in actual numerical calculations is how to construct $|n\rangle_C$ and $|\mu\rangle^{\mathrm{sub}}_{NC}$.
This method can be obtained by using Eq. (\ref{eq:TB}).
At first, consider a norm kernel,
\begin{equation}
  I_{NC}(q'_1, q'_2) :=\langle q ; q - q'_1|q;q - q_2\rangle
  = {}_{NC}\langle q'_1 ; \psi|q'_2; \psi\rangle_{NC}
  {}_C\langle q|q\rangle_C.
\end{equation}
Since ${}_C\langle q|q\rangle_C$ does not depend on $q$, $I_{NC}(q'_1, q'_2)$ is independent of $q$.
Conversely, the $q$ dependency means that one has failed to construct the separable collective subspace, except by accidental success.
Diagonalizing $ I_{NC}$ yields the orthonormal states,
\begin{equation}
  |\mu\rangle^{\mathrm{sub}}_{NC} :=\int_{\mathcal{M}_{C}} dq'\,f_{NC}^{\mu}(q')|q' ; \psi\rangle_{NC}.
\end{equation}
Next, consider the orthonormal set $\{|m\rangle_C\}_m$, but $\{|q\rangle_C\}_q$ are the eigenstates of $\hat{Q}_C$ and are already orthonormalized.
Thus, using arbitrary orthonormalized functions $\{f_C^{m}(q)\}_m$, one can define it as
\begin{equation}
  |m\rangle_C := \int_{\mathbb{R}} dq\, f_C^{m}(q)|q\rangle_C.
\end{equation}
Then, $|m, \mu\rangle$ can be evaluated as
\begin{equation}
  |m, \mu\rangle = \int_{\mathbb{R}} dq \int_{\mathcal{M}_C} dq'\,f_C^m(q) f_{NC}^{\mu}(q')|q ; q'\rangle.
\end{equation}
One may perform the numerical calculation for the DGCM straightforwardly with the dynamical path $|q,p\rangle$.
Then, the complete orthonormal set is derived as
\begin{gather}
  I_{NC}(q'_1, q'_2)
  = \int_{\mathbb{R}} \frac{dp}{2\pi} \langle q - q'_1|q - q_2, p\rangle e^{-ipq}
  \times \mathrm{const.}, \\
  |m, \mu\rangle = \mathrm{normalization}\times\int_{\mathbb{R}} \frac{dp}{2\pi}\int_{\mathcal{M}_C} dq'\, \tilde{f}^{m}_C(p)f^{\mu}_{NC}(q')|q', p\rangle,
\end{gather}
where $\{\tilde{f}^{m}_C(p)\}_m$ are arbitrary orthogonal functions.

\section{Summary and future perspectives}
I have discussed the structure of the collective subspace for the dynamical GCM, which is an improved GCM with the symplectic form in analytical mechanics.
Then, the reasonable boundary condition for the conjugation condition and some assumptions yield the locally separable collective subspace (\ref{eq:TPS}) for general collective motions.
It has also been shown that using the conjugate momentum operator restricts the collective subspace to the simple tensor product of the collective part and the non-collective part (\ref{eq:SHS}).
Then, even if one does not know the specific form of the conjugate momentum operator, in the case of the vibrational motions, the separable collective subspace can be obtained without it by the constrained variational method (\ref{eq:CVE1}).
This approach is numerically computable and practical.
In addition, I have shown that the collective subspace is invariant under the local symplectic transformation of the collective operator compatible with the conjugation condition.
Thus, the local operators are not always relevant to describing the collective motion in the context of the DGCM.
Furthermore, this analytical method can be immediately applied to the structural analysis of the VAP function space and has revealed that under the specific condition, the space has the untwisted structure (\ref{eq:UTV}).
This untwisted VAP is useful in considering the separable collective subspace.

In this paper, the existence of the collective operator has been assumed.
It cannot be overemphasized that identifying it is difficult, except in trivial cases such as the symmetry operator, and empirically chosen operators are often used in practical applications.
Then, whether the motion associated with the assumed operator is collective should be carefully investigated, and the DGCM is one of the frameworks which can be used for that purpose.
Therefore, it is a meaningful task to reconsider the conventional GCM calculations from the DGCM.
In particular, in the large-amplitude collective motions such as nuclear fission, the simple RPA is not valid, so the DGCM becomes essential.
One of the important questions is whether multipole operators can yield the proper collective coordinates for describing nuclear fission.

In this paper, the collective motion is characterized by Eq. (\ref{eq:NIH}).
However, there is no guarantee that any given motion that one can recognize as a collective motion has the separable form.
Instead, there may be a collective motion in which quantum entanglement between subsystems $C$ (collective part) and $NC$ (non-collective part) is intrinsically significant.
In order to deal with this problem, it is helpful to systematically investigate what kind of Hamiltonian is approximated well with a given collective subspace that does not have the simple tensor product structure.

\section*{Acknowledgment}
% The authors thank to Dr. AAAAAA and Dr. BBBBBB for fruitful discussions.
% This work was supported by JSPS KAKENHI Grant Nos. 21J22348, XXXXXXXX, and YYYYYYYY.
The author thanks K. Hagino and K. Yoshida for fruitful discussions.
This work was supported by JSPS KAKENHI Grant No. 21J22348.


%\bibliographystyle{ptephy}
%\bibliography{sample}
%
% once the .bbl file has been generated then place the text in your article.

\vspace{0.2cm}
\noindent

%This is added by T. Yoneya (editor-in-chief) on 2020/07/09.

% \let\doi\relax

%without this code before the command "\begin{thebibliography}{}" , an error will be %flagged. When the bibliography is provided as separate .bib file, then this code %should be placed above the commands "\bibliographystyle{}" and "\bibliography{}" %inside the main TeX file.

% \bibliographystyle{apsrev4-2}
% \bibliographystyle{jplain}
\bibliographystyle{ptephy}
\bibliography{ref}

\begin{thebibliography}{10}

\bibitem{Ring_Schuck}
P.~Ring and P.~Schuck,
\newblock {\em The Nuclear Many-Body Problem},
\newblock  (Springer-Verlag, New York, 1980).

\bibitem{Gogny}
J.~Decharg\'e and D.~Gogny, Phys. Rev. C, {\bf 21}, 1568--1593 (Apr 1980).

\bibitem{Skyrme}
T.~H.~R. Skyrme, Philos. Mag., {\bf 1}(11), 1043--1054 (1956).

\bibitem{RG84}
Erich Runge and E.~K.~U. Gross, Phys. Rev. Lett., {\bf 52}, 997--1000 (Mar
  1984).

\bibitem{HW53}
David~Lawrence Hill and John~Archibald Wheeler, Phys. Rev., {\bf 89},
  1102--1145 (Mar 1953).

\bibitem{FV76}
H.~Flocard and D.~Vautherin, Nucl. Phys. A, {\bf 264}, 197--220 (1976).

\bibitem{BD90}
P.~Bonche, J.~Dobaczewski, H.~Flocard, P.-H. Heenen, and J.~Meyer, Nucl. Phys.
  A, {\bf 510}(3), 466 -- 502 (1990).

\bibitem{Bender2003}
Michael Bender, Paul-Henri Heenen, and Paul-Gerhard Reinhard, Rev. Mod. Phys.,
  {\bf 75}, 121--180 (Jan 2003).

\bibitem{NV11}
T.~Nikšić, D.~Vretenar, and P.~Ring, Prog. Part. Nucl. Phys., {\bf 66}(3),
  519 -- 548 (2011).

\bibitem{E16}
J~Luis Egido, Phys. Scr., {\bf 91}(7), 073003 (Jun 2016).

\bibitem{RR18}
L~M Robledo, T~R Rodr{\'{\i}}guez, and R~R Rodr{\'{\i}}guez-Guzm{\'{a}}n, J.
  Phys. G: Nucl. Part. Phys., {\bf 46}(1), 013001 (Dec 2019).

\bibitem{SO06}
Satoshi Shinohara, Hirofumi Ohta, Takashi Nakatsukasa, and Kazuhiro Yabana,
  Phys. Rev. C, {\bf 74}, 054315 (Nov 2006).

\bibitem{BH08}
Michael Bender and Paul-Henri Heenen, Phys. Rev. C, {\bf 78}, 024309 (Aug
  2008).

\bibitem{YM10}
J.~M. Yao, J.~Meng, P.~Ring, and D.~Vretenar, Phys. Rev. C, {\bf 81}, 044311
  (Apr 2010).

\bibitem{TR11}
Tomás~R. Rodríguez and J.~Luis Egido, Phys. Lett. B, {\bf 705}(3), 255 -- 259
  (2011).

\bibitem{RE11}
Tom\'as~R. Rodr\'{\i}guez and J.~Luis Egido, Phys. Rev. C, {\bf 84}, 051307
  (Nov 2011).

\bibitem{YB13}
J.~M. Yao, M.~Bender, and P.-H. Heenen, Phys. Rev. C, {\bf 87}, 034322 (Mar
  2013).

\bibitem{FS13}
Y.~Fukuoka, S.~Shinohara, Y.~Funaki, T.~Nakatsukasa, and K.~Yabana, Phys. Rev.
  C, {\bf 88}, 014321 (Jul 2013).

\bibitem{BA14}
B.~Bally, B.~Avez, M.~Bender, and P.-H. Heenen, Phys. Rev. Lett., {\bf 113},
  162501 (Oct 2014).

\bibitem{Yao2014}
J.~M. Yao, K.~Hagino, Z.~P. Li, J.~Meng, and P.~Ring, Phys. Rev. C, {\bf 89},
  054306 (May 2014).

\bibitem{YZ15}
J.~M. Yao, E.~F. Zhou, and Z.~P. Li, Phys. Rev. C, {\bf 92}, 041304 (Oct 2015).

\bibitem{EJ20}
J.~Luis Egido and Andrea Jungclaus, Phys. Rev. Lett., {\bf 125}, 192504 (Nov
  2020).

\bibitem{PT62}
P.E. Peierls and D.J. Thouless, Nucl. Phys., {\bf 38}, 154 -- 176" (1962).

\bibitem{BR15}
Marta Borrajo, Tomás~R. Rodríguez, and J.~{Luis Egido}, Phys. Lett. B, {\bf
  746}, 341 -- 346 (2015).

\bibitem{EB16}
J.~Luis Egido, Marta Borrajo, and Tom\'as~R. Rodr\'{\i}guez, Phys. Rev. Lett.,
  {\bf 116}, 052502 (2016).

\bibitem{ST15}
Mitsuhiro Shimada, Shingo Tagami, and Yoshifumi~R. Shimizu, Prog. Theor. Exp.
  Phys., {\bf 2015}(6), 063D02 (2015).

\bibitem{Shimada2016}
Mitsuhiro Shimada, Shingo Tagami, and Yoshifumi~R. Shimizu, Phys. Rev. C, {\bf
  93}, 044317 (Apr 2016).

\bibitem{Ushitani2019}
Masaki Ushitani, Shingo Tagami, and Yoshifumi~R. Shimizu, Phys. Rev. C, {\bf
  99}, 064328 (Jun 2019).

\bibitem{GR80}
K.~Goeke and P.-G Reinhard, Annu. Phys. (N. Y.), {\bf 124}(2), 249 -- 289
  (1980).

\bibitem{MM80}
Toshio Marumori, Toshihide Maskawa, Fumihiko Sakata, and Atsushi Kuriyama,
  Prog. Theor. Phys., {\bf 64}(4), 1294--1314 (Oct 1980).

\bibitem{M84}
Masayuki Matsuo, Prog. Theor. Phys., {\bf 76}(2), 372--386 (Aug 1986).

\bibitem{MN00}
Masayuki Matsuo, Takashi Nakatsukasa, and Kenichi Matsuyanagi, Prog. Theor.
  Phys., {\bf 103}(5), 959--979 (May 2000).

\bibitem{NM06}
Takashi Nakatsukasa, Kenichi Matsuyanagi, Masayuki Matsuo, and Kazuhiro Yabana,
  Rev. Mod. Phys., {\bf 88}, 045004 (Nov 2016).

\bibitem{KE08}
P.~Klüpfel, J.~Erler, P.~G. Reinhard, and J.~A. Maruhn, Eur. Phys. J. A, {\bf
  37}(3), 343--355 (2008).

\bibitem{HH21}
N.~Hizawa, K.~Hagino, and K.~Yoshida, Phys. Rev. C, {\bf 103}, 034313 (Mar
  2021).

\bibitem{HH22}
N.~Hizawa, K.~Hagino, and K.~Yoshida (2022),  {{arXiv:2204.01995}}.

\bibitem{jan64}
B.~Jancovici and D.H. Schiff, Nucl. Phys., {\bf 58}, 678 -- 686 (1964).

\bibitem{BW68}
D.M. Brink and A.~Weiguny, Nucl. Phys. A, {\bf 120}(1), 59 -- 93 (1968).

\bibitem{KK12}
Yoshiko Kanada-En'yo, Masaaki Kimura, and Akira Ono, Prog. Theor. Exp. Phys.,
  {\bf 2012}(1) (Aug 2012).

\bibitem{PV80}
J.~P. Provost and G.~Vallee, Commun. Math. Phys., {\bf 76}(3), 289--301 (Sep
  1980).

\bibitem{QGT}
F~Wilczek and A~Shapere,
\newblock {\em Geometric Phases in Physics},
\newblock  (WORLD SCIENTIFIC, 1989).

\bibitem{GB20}
A.~Gianfrate, O.~Bleu, L.~Dominici, V.~Ardizzone, M.~De~Giorgi, D.~Ballarini,
  G.~Lerario, K.~W. West, L.~N. Pfeiffer, D.~D. Solnyshkov, D.~Sanvitto, and
  G.~Malpuech, Nature, {\bf 578}(7795), 381--385 (Feb 2020).

\bibitem{LG21}
Diego Liska and Vladimir Gritsev, SciPost Phys., {\bf 10}, 20 (2021).

\bibitem{Sheikh2002}
J.~A. Sheikh, P.~Ring, E.~Lopes, and R.~Rossignoli, Phys. Rev. C, {\bf 66},
  044318 (Oct 2002).

\bibitem{K98}
Y.~Kanada-En'yo, Phys. Rev. Lett., {\bf 81}, 5291--5293 (Dec 1998).

\bibitem{RS04}
R.~R. Rodríguez-Guzmán and K.~W. Schmid, Eur. Phys. J. A, {\bf 19}(1), 45--59
  (2004).

\bibitem{EC10}
J.~Eisert, M.~Cramer, and M.~B. Plenio, Rev. Mod. Phys., {\bf 82}, 277--306
  (Feb 2010).

\end{thebibliography}

\end{document}